# Towards a participatory E-learning 2.0

A new E-learning focused on learners and validation of the content

Boubker Sbihi
Laboratory LIROSA
Faculty of Sciences
Tétouan, Morocco
Bsbihi@hotmail.com

Kamal Eddine El Kadiri
Laboratory LIROSA
Faculty of Sciences
Tétouan, Morocco
Elkadiri@uae.ma

*Abstract*—Our aim is to propose a collaborative methodological approach centred on learners and based on the Web 2.0 tools in order to make E-learning 2.0. It is based on a process consisting of four iterative steps which are: grouping, collaborating, validating and publishing content. In this context, learners will be the creators of the content of assigned courses in a virtual meeting through the chat. These contents will be validated after a pedagogical monitoring by the instructor through the class's blog and merged into a single course content published on a class' wiki. Social interaction and sharing of files on the web will be the responsibility of social networks. The rest of the web 2.0 tools such as RSS feeds, tags, podcasts and video casts will be used as complementary tools in order to improve the quality of training. This methodological approach will allow E-learning 2.0 by ensuring a better interactivity, collaboration, sharing and an optimal exploitation of collaborative intelligence in the classroom.

## I. INTRODUCTION

With the advent of the 21st century, the Internet World Wide Web saw a real evolution in containers and contents. Indeed, taking into consideration containers, the Internet has become more mobile, faster and more secure with the advent of wireless networks and very high-speed rate. For content, it has become a mine of ideas and a library of information thanks to the important masses of data resulting from the collaboration of millions of users in the context of Web 2.0 that has changed the place of knowledge in our societies. It has contributed massively to the economic and social development of countries in the world and especially those in process of development, as it is the case of African countries [2].

Moreover, the educational and training sector has benefited from this situation to develop distance education in order to improve the quality of the educational system and expand access to training.

Training has been proposed to all kinds of learners as part of qualifying courses and continuing education for personal and professional needs. Its success and large diffusion are due to the emergence of a powerful range of LMS platforms, especially those under a free license, which offer various functionalities of pedagogy and training [5].

In this context, universities have also benefited from these new forms of transmission of knowledge by transforming the current teaching from a package of information to be delivered, to a permanent open learning environment which has given birth to the digital campus.

In the professional environment, E-learning has played a very important role in the life of companies by increasing productivity and reducing costs and time for employee training. Indeed, it provides great flexibility by allowing learners to choose the specified place and time that they wish to devote to their training according to their professional and personal availability. This has made it possible for the staff to implement great projects at lower costs and thus bolster the company in a world of competitiveness in the context of globalization. E-learning has also given place to an even better third form of education which attempts to maximize the advantages of face-to-face discussion and online methods as two balanced and combined teaching modes, aiming at the development of the learner's knowledge [15].

In addition, in 2004, Web 2.0 appeared as a new vision of the Web which considered the user not a simple consumer of information but as a potential producer of the web content [16]. This radical change has significantly increased the quantity of information and allowed a certain organization of users in the form of communities participating in the production, communication, sharing and diffusion of content.

In this context, the Web has become a platform of tools that each user can utilize, and thereby increase the utility of the network by allowing the existence of a collective intelligence. Although Web 2.0 tools are easily used for social purposes, they could also be used by learners to complete learning experiences [14].

Until now, all these new possibilities have been exploited to create informal learning at a lower cost; for example, the learning through blogs [7] [8], podcasts and videocasts [20] , or making hybrid education using blogs and podcast [17]. However, Web 2.0 does not offer a specific tool for E-learning similar to those used for bookmarks' sharing, multimedia's files sharing, collaborative tools or those available for Office Online.

The aim of our article is to benefit to the maximum from the Web 2.0 participative approach by adding the concept of validation of content [22] to define a new method of collaborative learning; E-learning 2.0. For this, we propose the elements of a collaborative methodological approach based on the Web 2.0 tools which will put the learner at the centre of the learning, and will be based on his/her participation to animate and feed the contents of the courses in a context of collaborative distance learning.

In the following paragraph, we present the concepts of classical E-learning; in paragraph 3, we expose the concepts and tools of Web 2.0; and then, in paragraph 4, we propose the concept of E-learning 2.0 before introducing elements of our approach and methodology based and centred on the learner who will make possible collaborative E-learning by using the tools of Web 2.0. The last paragraph will present a general conclusion putting forward a series of perspectives.

## II. THE CLASSIC E-LEARNING

E-learning is a type of teaching supported by electronic means which comes from the broad diffusion of information through the networks and channels of telecommunication in education. It is defined by the American Society for Training & Development [9] as the use of the Internet and digital technologies to create experiences of participating in education. The European Commission defines it as the use of new multimedia technologies and the Internet to improve the quality of education and training through distance access to resources and services, collaboration and exchanges [4]. It can be seen through several points of view such as the economic, organizational, educational and technological aspects. Moreover, the E-learning allows a great flexibility in promoting and facilitating learning at the sought place, time and pace. This flexibility in the allocation of time and personal rhythm of the courses has led to good results on the level of teaching [21]. It also reduces the time and training costs and facilitates the exchange of information.

In a similar way to the traditional teaching, this type of teaching has four actors who play specific roles within the process of learning and knowledge acquisition. The principal actors of classic E-learning and their roles are grouped in the following table:

TABLE I. ACTORS OF THE CLASSIC E-LEARNING

| Actor | Role |
|---|---|
| Teacher | Produces the courses and exercises |
| Learner | Takes courses, does homework |
| Tutor | Registrates, monitors and supports learners |

To carry out E-learning, we often use a platform for learning LMS (Learning management system) which is defined by Prat as a management learning program which assists the conduct of E-learning such as management training, courses, learners, virtual classrooms and statistics [19]. It is a virtual learning environment that provides tools necessary for principal users to communicate, organize, produce and share [19]. To integrate the collaborative approach in E-learning, the LMS has migrated to LCMS (Learning Content Management System) which are content management systems of CMS (Content Management System) to which the functionalities of a LMS are added. They are focused on integration, collaboration, diffusion and management of content to create, validate, publish and manage the learning content as part of an E-learning.

In this context and in order to describe the learning scenarios that represent the description of the learning situation, standards and educational modeling languages emerged for the purpose standardization and reuse. At the level of the standards, the community of research developed until now several specifications and standards such as AICC, IMS, LOM or the SCORM [3], aiming all at promoting the reuse of the teaching resources related mainly to the educational contents. Let us note well that SCORM attempts to synthesize different systems (AICC, IMS, LOM) in order to make it easily usable standard, enabling the reuse of resources.

Concerning the educational modeling languages that are proposed as a solution by the scientific community to model the learning activities and promote the use of certain types of teaching models in order to define learning scenarios [10], we can cite for example, CPM's (Cooperative Problem-based learning Metamodel)scripting language, or language LDL (Learning Design Language) or the PoEML language (Perspective-oriented Educational Modeling Language) [10]. They allow formalizing, standardizing and reusing information coming from various systems on education. Some experiments show that the best solutions to facilitate the reuse of e-learning tools are based on collaboration and practice. The use of web technologies in e-learning is further reinforced with Web 2.0 which proposes a set of simple, easy to use tools that promote social interaction between internet users, in a context where everyone is able to add and modify information [1]. The idea of integrating them into an LMS platform is the objective of the future platforms. The first in the range of open source that started this integration is the MOODLE platform (Modular Object-Oriented Dynamic Learning Environment) [13] which already integrates the tools of knowledge management issue of Web 2.0 such as wiki, RSS feeds or blogs. These tools support the collaborative work of a community centred on a learning project. The difficulty arising is that all users must control the use of the platform; which requires installation and expensive maintenance.

## III. THE WEB 2.0

Since 2004, the Web 2.0 [16] has been presented as a new vision of the web which places the user at the centre of the web and regards him/her as a potential producer of content and therefore a vital actor for the web. Indeed, it is based on a participating and collaborative process that allows any user to produce, communicate, share and diffuse content easily while personalizing its navigation.

In this context, each user will increase the utility of the network and will participate in the creation of a collective intelligence and thus becomes part of a participating community, passing from the authority to popularity and from relevance to influence. This fundamental change has created new relations with the user, and is presented as a real evolution of the web which has increased the quantity of information while offering to the user greater interactivity, accessibility and has facilitated his/her search for information and possibilities of publications.

According to Devis, web 2.0 is a social philosophy that aims at abandoning the individual control over matters and gaining a greater number of participation [6]. Miller defines the web 2.0 as an equal regrouping of the evolution and revolution which benefitted from existing standards such as HTML, CSS and XML and the Web browser [12]. According to Richard Mac Manus, Web 2.0 is social and open; it leaves the control of data and combines the global with the local. It corresponds to new manners to search and access to content. It is a ready platform for educators, media, politics, communities and for everyone else.

Richard Mac Manus says that definitions may come better from engineers, sociologists, designers, philosophers, educators, businessmen. Yet all of them count [11]. There is no precise definition of Web 2.0 and this concept is composed of as many definitions as users.

Concretely, Web 2.0 has three dimensions: social, technical and economic. On the social side, Web 2.0 is presented as a real network of social interaction based on the participation of users. In other words, anyone can easily create an information space accessible to anyone, anywhere, and which allows anyone to put any content on line. At the technical level, Web 2.0 makes access easy to the production and use of information and documentation through the changeover from software installation to the exploitation of online services by the use of the AJAX, CSS, XML.

At the economic level, the financing of the Web 2.0 sites occurs primarily through the commercial publicities, offers and traffics' networks instead of gifts or payments for licenses to use proprietary software. Web 2.0 presents seven concepts According to Tim O'Reilly [16] which are:

1-The Web passes from a collection of websites to a complete platform of services and Web applications to users.

2 - Users become producers and applications co-developers, thus passing from the concept of "software product" to "software service".

3. The service improves when the number of users increases and the users of Web 2.0 have unique data and difficult to recreate. The resources of the web increase with the increase in the number of users participating to create a collective intelligence

4. The riches is in the data which benefit from the open source movement opposing gradually to the universe proprietary owners.

5. Create collective intelligence from the collaboration of the users is the key factor to profits on the market

6. Create flexible interfaces based on the standards and web protocols (AJAX, CSS, XML)

7. Open Web on various hardware devices and mobile giving the opportunity to release software to the PC.

The new tools of Web 2.0 focusing on users permit an evolution in content, and change the manner in which the user utilizes the web. Web 2.0 has a range of tools such as blogs, wikis or social networks that attract a large number of Internet users around the world. In addition to these three classes of basic tools, other complementary tools are added such as:

• RSS feeds which make it possible to follow all the news on a specific field and be notified by email on the latest news from a site,

• Tags that enable the description of keywords by the user contrary to the webmaster, in this context, the contents are classified by users.

• The platform of multimedia files sharing as podcasts and video cast, which are deposits of files in the form of audio and video, can be downloaded anytime, anywhere and by anyone without requiring installing any software.

The principal tools of Web 2.0 are six: blogs, wikis, social networks, RSS feeds, tags, podcast and videocast presented in the following table:

TABLE II. WEB 2.0'S TOOLS

| Tool | Utility |
| --- | --- |
| Blog | Publication and comment the information |
| Wiki | Collective intelligence |
| Social network | Creation of online communities |
| RSS feed | Regular informational monitoring |
| Tag | Improve and Personalize the research |
| Podcast and Videocast | Sharing audio and video files |

As the table shows, the tools of web 2.0 are very rich and powerful. They do not require installation and they are largely popular especially among younger generations. They propose a learning environment by offering a favorable context to exchange and create social interactions. However Web 2.0 does not offer specific tools to make E-learning. That's why our research aims at proposing a methodology to do it.

IV. THE CLASSIC E-LEARNING

The classic E-Learning consists of the 'use of the tools and platforms like LMS and LCMS to learn and make an E-learning. It was a great success thanks to a broad range of special offers especially in open source. However, LMS have limitations such as their slowness and their difficult use which requires some special knowledge. The content was the second concern of those responsible for E-learning. This is why good platforms can be found but with very limited or empty content in some cases. In this context, the learner is a passive spectator as it is the case for users of the web 1.x. Web 2.0 has radically changed the techniques to online information access through the various synchronous and asynchronous tools. It puts users at the centre of the network and provides opportunities for the creation of decentralized learning communities. With its appearance, a real interactivity based on the collaboration of the word users was born, and has therefore given new paradigms 2.0 such as the library 2.0, enterprise 2.0, .... the field of teaching does not come out of this rule, and like the web migrated from version 1.0 to 2.0 with the movement of 2.0, the E-Learning migrated to the E-learning 2.0. It stems from the blending of classic E-Learning with the Web 2.0 tools and services which allow the emergence of a new model of learning as shown in the following figure:

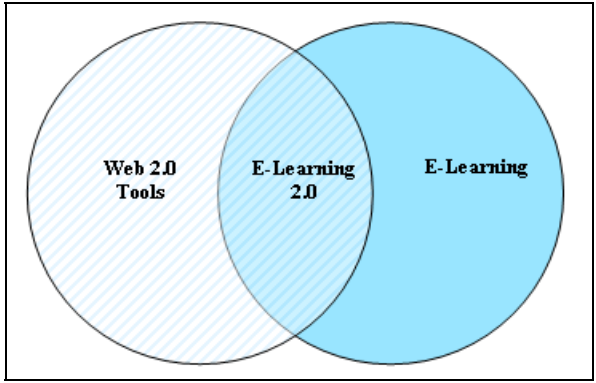

Figure 1. Origin of the E-Learning 2.0

E-learning 2.0 is closer to a social network and a community of practice articulated around a field of interest, where members interact and learn together [7]. This is a new mode of learning based on Web 2.0 which allows restoring power to the user and creates the dynamic horizontal community (learner-learner or teacher-teacher) and vertical community (teacher-learner), thus migrating from transmissive unidirectional media to a collaborative learning.

The origin of the integration of the collaborative aspect in teaching is an old concept defined since 1968 by Piaget [18], founder of the constructivist approach to learning, which says that knowledge is constructed by the learner and not transmitted just as it is by the teacher. You can find many different definitions of E-Learning 2.0 as it is the case for Web 2.0. We define the E-Learning 2.0 as a new environment for E-learning that places the learner at the centre of the training through the tools of Web 2.0. It does allow passing from transmission to collaboration. This mode of learning is approved by the Web 2.0 tools which are:

• Collaborative, easy and simple to use;

• Allow to be used at the world level, especially by younger generations;

• Allow a real interactivity;

• Allow developing practices, sharing of educational content, teaching methods;

• The majority of these tools are free and without publicity.

E-Learning 2.0 has introduced the concept of a learning community which focuses on supporting the development and solving educational problems through online collaboration. It allows communication among learners and promotes the creation of a collective intelligence, recreating the traditional learning of the classroom and interactions between learners. It is learner-centred contrary to teacher-centred and is based on the concept of collaboration using Web 2.0 tools to carry out the contents of the training. It is mainly characterized by the interaction, collaboration, autonomy and responsibility. The following table gives the differences between the E-Learning 2.0 and E-Learning 1.0:

TABLE III. DIFFERENCE BETWEEN E-LEARNING 1.0 AND E-LEARNING 2.0

| *E-Learning 1.0* | *E-Learning 2.0* |
|---|---|
| Platform LMS and LCMS | Tools Web 2.0 |
| Based on teacher | Based on learner |
| The teacher produces | The teacher validates |
| Learner is a spectator | Learner is a producer |
| Exchange with the class | Exchange with the community |

In this context, learners have more freedom, more responsibility and take control of their training while creating their learning environment. They will be able to communicate with other learners and experts outside the boundaries of classrooms. They can work remotely through a shared whiteboard, make chats through their webcams, evaluate and comment on the learning contents collectively, publish using blogs, produce a collaborative document through wikis, and facilitate the redistribution of other sites through the tags and be alerted when there's a new content using the RSS feed. This situation allows an opening, makes the experience more interactive, enriches the debates and increases the critical sources. In this context, we can mention four possibilities to make collaborative E-learning which are:

• The addition of Web 2.0 tools to a distance platform as it has been done for the MOODLE platform which has begun to integrate blogs. However, the burden of this platform, the difficulty of its use by the public, in addition to the fact it is not the only one to be used, are major limitations.

• The educational portal which proposes to the teachers to share and exchange freely their educational resources. It allows to discuss on a forum and to write collaboratively news' teaching via a blog. The problem with this initiative is the lack of participation and also the difference in participation between teachers.

• Use one of Web 2.0 tools such as blogs or social networks as popular tools to make an E-learning 2.0. Indeed, the most used tools are blogs which encourage the development of learner autonomy or social networks that are used to promote effective communication and collaborative work between teachers and students.

• Group Web 2.0 tools in an E-learning collaborative approach. We propose in this research a methodology for E-learning based on Web 2.0 tools. This method has three objectives which are:

Teaching richness (online platform courses, exercises.);

Social Interaction (in classroom and online);

Cost-effectiveness.

### V. TOWARDS A METHOLOGY FOR THE E-LEARNING

The principal actors of classic E-learning are: the teacher who produces the course, the learner who follows the courses, the tutor who is responsible for monitoring and assisting learners, and the administrator who is responsible for technical problems. In this new vision, the actors will remain the same but their roles will change. The administrator, however, will no longer be needed as an actor since the Web 2.0 tools are easy to use and to parameterize by any user without any computer knowledge. However, the information produced by learners cannot be considered in an E-learning without validation from the instructor. After the corrections, the instructor may give learners the opportunity to circulate good information thus privileging collaborative work.

In this context, three roles are assigned to actors: learning, teaching and tutoring. We propose to maximize the production of the learner and to give the possibility to the teacher to validate the contents. The tutor will be a guide and a publisher of valid contents. The following table shows the role of E-Learning 2.0:

TABLE IV. ROLES OF ACTORS OF THE E-LEARNING 2.0

| Actors | Role |
|---|---|
| Learner | Producing contents |
| Teacher | Validating of contents |
| Tutor | Monitoring of the publications |

Each Web 2.0 tool can be used to maximize the profitability and quality of learning and teaching. The following table shows the role of Web 2.0 tools in an E-learning collaborative environment:

TABLE V. ROLES OF THE E-LEARNING 2.0 TOOLS

| Social Tool | Role |
|---|---|
| Blog | publication of courses, pedagogical coaching |
| Wiki | Creation of common documents |
| Forum | Exchange between learners for resolve a problem |
| Social network | Forming learning communities |
| Tags | Improving research by involving the learner |
| Feed RSS | Improving the dissemination by notifying the actor |
| Podcast | Learning by listening the audio files |
| Videocast | Learning by watching the video files |

In this context, we can subdivide the web 2.0 tools into two categories: primary and secondary tools.

• The primary tools: are the basic tools of our E-learning 2.0 methodological approach based on the Web 2.0 tools and are used by an actor directly.

• The secondary tools: are the proposed complementary tools to the primary tools aiming at perfecting our E-learning 2.0 methodological approach. They can be used with one or more primary tools. The following table lists the Web 2.0 tools in two types:

TABLE VI. CLASSIFICATION OF THE E-LEARNING 2.0 TOOLS

| Primary tools | Secondary tools |
|---|---|
| Blog | Tag |
| Wiki | Feed RSS |
| Forum | Podcast |
| Social network | Videocast |

In order to propose rich contents blending various media, we propose a methodological approach which consists of four iterative steps. The general architecture and the four steps are illustrated in the following figure:

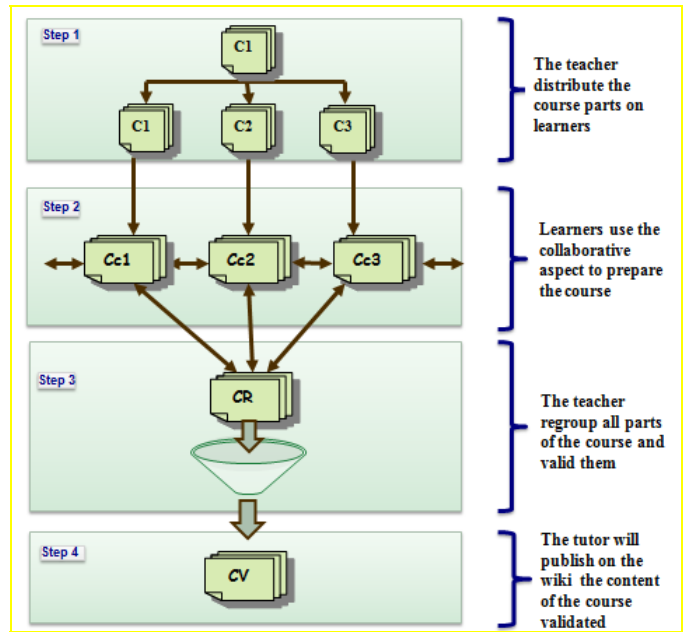

Figure 2. Steps of our methodology

Concerning the tools used to assemble this methodological approach, certain primary tools will be used in one or the totality of the four steps while the secondary tools can be used in all steps. In step 1, some tools will be created and used in step 2, then used for final validation in step 3, and published in step 4. The following table shows the number of tools used in the four steps.

Teaching richness (online platform courses, exercises.);

Social Interaction (in classroom and online);

Cost-effectiveness.

### V. TOWARDS A METHOLOGY FOR THE E-LEARNING

The principal actors of classic E-learning are: the teacher who produces the course, the learner who follows the courses, the tutor who is responsible for monitoring and assisting learners, and the administrator who is responsible for technical problems. In this new vision, the actors will remain the same but their roles will change. The administrator, however, will no longer be needed as an actor since the Web 2.0 tools are easy to use and to parameterize by any user without any computer knowledge. However, the information produced by learners cannot be considered in an E-learning without validation from the instructor. After the corrections, the instructor may give learners the opportunity to circulate good information thus privileging collaborative work.

In this context, three roles are assigned to actors: learning, teaching and tutoring. We propose to maximize the production of the learner and to give the possibility to the teacher to validate the contents. The tutor will be a guide and a publisher of valid contents. The following table shows the role of E-Learning 2.0:

TABLE IV. ROLES OF ACTORS OF THE E-LEARNING 2.0

| Actors | Role |
|---|---|
| Learner | Producing contents |
| Teacher | Validating of contents |
| Tutor | Monitoring of the publications |

Each Web 2.0 tool can be used to maximize the profitability and quality of learning and teaching. The following table shows the role of Web 2.0 tools in an E-learning collaborative environment:

TABLE V. ROLES OF THE E-LEARNING 2.0 TOOLS

| Social Tool | Role |
|---|---|
| Blog | publication of courses, pedagogical coaching |
| Wiki | Creation of common documents |
| Forum | Exchange between learners for resolve a problem |
| Social network | Forming learning communities |
| Tags | Improving research by involving the learner |
| Feed RSS | Improving the dissemination by notifying the actor |
| Podcast | Learning by listening the audio files |
| Videocast | Learning by watching the video files |

In this context, we can subdivide the web 2.0 tools into two categories: primary and secondary tools.

• The primary tools: are the basic tools of our E-learning 2.0 methodological approach based on the Web 2.0 tools and are used by an actor directly.

• The secondary tools: are the proposed complementary tools to the primary tools aiming at perfecting our E-learning 2.0 methodological approach. They can be used with one or more primary tools. The following table lists the Web 2.0 tools in two types:

TABLE VI. CLASSIFICATION OF THE E-LEARNING 2.0 TOOLS

| Primary tools | Secondary tools |
|---|---|
| Blog | Tag |
| Wiki | Feed RSS |
| Forum | Podcast |
| Social network | Videocast |

In order to propose rich contents blending various media, we propose a methodological approach which consists of four iterative steps. The general architecture and the four steps are illustrated in the following figure:

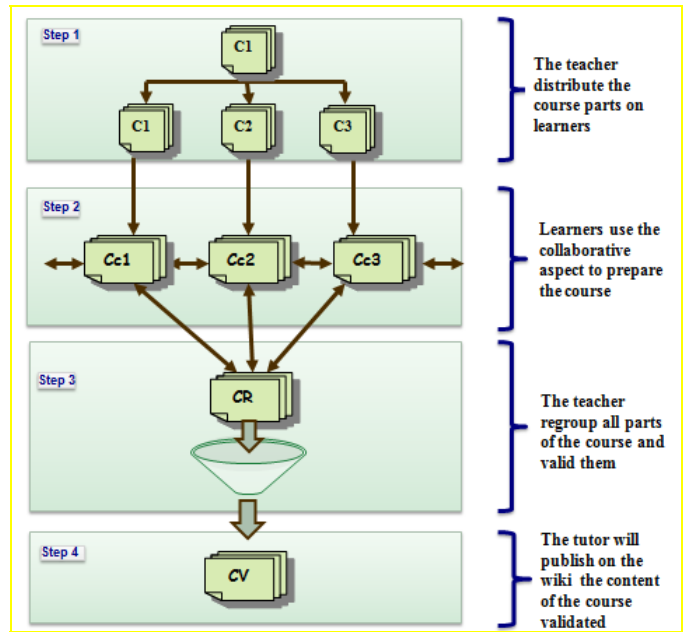

Figure 2. Steps of our methodology

Concerning the tools used to assemble this methodological approach, certain primary tools will be used in one or the totality of the four steps while the secondary tools can be used in all steps. In step 1, some tools will be created and used in step 2, then used for final validation in step 3, and published in step 4. The following table shows the number of tools used in the four steps.

TABLE VII. USING THE WEB 2.0 TOOLS IN THE FOUR STAGES OF THE METHODOLOGY

| Actors | Primary tools | Steps | | | |
|---|---|---|---|---|---|
| | | *Step 1* | *Step 2* | *Step 3* | *Step 4* |
| | operation | Creation | Use | Validation | Publication |
| Learner | Blog | 2 | Yes | Yes | - |
| | Wiki | 0 | - | - | - |
| | Forum | 1 | Yes | Yes | - |
| | Social network | 1 | Yes | Yes | - |
| Teacher | Blog | 1 | Yes | Yes | - |
| | Wiki | 0 | - | - | - |
| | Forum | 0 | - | - | - |
| | Social network | 1 | Yes | Yes | - |
| Tutor | Blog | 1 | Yes | Yes | - |
| | Wiki | 1 | - | - | Yes |
| | Forum | 0 | - | - | - |
| | Social network | 1 | Yes | Yes | - |

As an example, we take the example of a 3 courses module. We will have a tutor for the module and three teachers. Each teacher will be responsible for a course for a group of 20 learners who will follow the three courses.

First, learners, teachers and tutors will meet online or in class to set the general objectives. In this context, 3 types of tools will be created:

- One class blog managed by the tutor to monitor the pedagogical follow-up: teaching, tutoring and research.

-Three professional blogs, one for the teachers, one for the learners and one for the tutors managed by someone of each type will be created to ensure the exchange of experiences and opinions.

- One Class forum administered by the tutor: It will be the place where problems will be exposed with proposed solutions by the E-learning actors.

- One social network of training that starts with a group of 20 learners, linked with a social network, will be able to reach a million users who can interact and share with external users.

- Two social networks, one for the teachers and another for the tutors, will be able to regroup many millions of users and professionals in the field. It can be the starting point to find other teachers who will like to adhere to the training.

- In the second step, the tools will be used to allow tutoring and exchanging.

- In the third stage, a final validation will be made by the teacher.

- In the fourth step, a final publication will be made by the tutor.

This approach will allow E-learning 2.0 guarantee a greater interactivity, collaboration, sharing and an optimal use of collaborative intelligence in the classroom at lower costs.

## VI. CONCLUSION

The aim of our proposal is to propose an iterative, collaborative and learner-centered methodological approach, based on the platform of Web 2.0 tools to make E-learning. In this context, learners will be the creators of courses' content, the teacher will be responsible for the coaching and final validation, and the tutor will be responsible for final publication. Social interactions at the level of the Web are possible by using social networks. This approach does not require many resources since the platform of Web 2.0 tools exist and many tools are free for use. It is just a question of giving each actor the adequate role.

It will enable non scientists to make and manage their training without being accompanied by a data processing specialist. Educational gains through the use of the collaborative approach will be possible and will encourage learners to open up on users' experiences while being guided by a teacher.

Among the sought perspectives, we can mention:

• Specific tools to make E-learning in the range of web 2.0 tools.

• Tools for blind people, which can expand further the learning community.

• Tools to organize brainstorming and create collaborative projects.

### AUTHORS PROFILE

Boubker Sbihi is pHD doctor and professor of computer science at the School of Information Science in morocco

Kamal Eddine El Kadiri is pHD doctor and professor of computer science at Faculty of Sciences of tétouan in Morocco.